\documentclass[12pt, preprint]{aastex}
\usepackage{graphicx}
\usepackage{epsfig}
\usepackage[nointegrals]{wasysym}

\usepackage{verbatim}

\begin{document}

\title{Statistical Tests of Chondrule Sorting}

\author{S. A. Teitler\footnote{Corresponding author: steitler@astro.wisc.edu}}
\affil{Department of Astronomy, University of Wisconsin--Madison, 475 N. Charter St., Madison, WI 53706, USA} 
\author{J. M. Paque}
\affil{Division of Geological and Planetary Sciences, California Institute of Technology, Pasadena, California 91125, USA}
\author{J. N. Cuzzi}
\affil{Space Science Division, NASA-Ames Research Center, Moffett Field, California 94035, USA}
\author{R. C. Hogan}
\affil{BAER, Inc., Sonoma, California 95476, USA}

\begin{abstract}

The variation in sizes of chondrules from one chondrite to the next is thought to be due to some sorting process in the early solar nebula. Hypotheses for the sorting process include chondrule sorting by mass and sorting by some aerodynamic mechanism; one such aerodynamic mechanism is the process of turbulent concentration (TC). We present the results of a series of statistical tests of chondrule data from several different chondrites. The data do not clearly distinguish between various options for the sorting parameter, but we find that the data are inconsistent with being drawn from lognormal or (three-parameter) Weibull distributions in chondrule radius. We also find that all but one of the chondrule data sets tested are consistent with being drawn from the TC distribution.

\end{abstract}


\section{INTRODUCTION}\label{intro}

Chondrites comprise the vast majority of meteoritic falls \citep{sd88}. Chondrules are the major component of chondrites, and were likely the most ubiquitous object in the solar nebula. An understanding of the formation conditions of chondrules will lead us to a better understanding of the early history of the solar system. It has been known for some time that chondrules vary in size between different classes of chondrites \citep{d76, h78a, r89, sh93}. This phenomenon has been interpreted as arising from some sorting process acting in the solar nebula between the time of chondrule formation and the time of chondrite parent body formation, or from some parent body process. A sorting process has also been proposed as a possible explanation for metal-silicate fractionation by acting on other components of chondrites, including metal and sulfide grains \citep{d76, sh93, sl93, has96, kmc97, bas98, sas98, as99, kmch99}. It is possible that the observed differences simply correspond to variations in the initial distributions of chondrules produced at different times and in different locations in the solar nebula, with no sorting or other post-formation alteration. The observed distribution of chondrules in any chondrite will reflect the initial distribution of chondrules produced, any sorting mechanisms that may have acted on the chondrules between formation and parent body assembly, and any parent body processes affecting the chondrule distribution. Chondrule distributions may vary between chondrite classes, or possibly even within chondrite classes.

Chondrule sizes in various samples have been described as approximately following lognormal \citep{kk79}, Rosin \citep{h78a, rg87} and Weibull distributions \citep{h78b, e96}. In general, a quantity $z$ with a lognormal distribution is described by a probability density 

\begin{equation}
f(z;\mu,\sigma) = \frac{1}{z\sigma\sqrt{2\pi}}{\rm exp}\left[-\frac{\left({\rm ln}(z)-\mu\right)^{2}}{2\sigma^{2}}\right], 
\end{equation}

\noindent where $\mu$ and $\sigma$ are the mean and standard deviation of ${\rm ln}(z)$, respectively. For a chondrule size distribution, $z$ is the chondrule radius $r$. Chondrule distributions in other parameters (e.g. mass, so that $z = \frac{4 pi}{3}\rho r^{3}$, with $\rho$ the chondrule density) can also be considered. The Weibull distribution comes in two-parameter and three-parameter varieties; the three-parameter variant has probability density 

\begin{equation}
f(z;\alpha,\beta,\gamma) = \frac{\beta}{\gamma}\frac{\left(z-\alpha\right)}{\gamma}^{\beta-1}{\rm exp}\left[-\left(\frac{z-\alpha}{\gamma}\right)^{\beta}\right]
\end{equation}

\noindent for $z > \alpha$. This distribution is characterized by a location parameter $\alpha$, shape parameter $\beta$, and scale parameter $\gamma$. The Weibull location parameter provides a minimum value for $z$, and corresponds to a constant additive shift in all data values, hence to a shift in the median, mean and mode. The shape parameter characterizes the shape of the distribution, and the scale parameter is a measure of the statistical dispersion of the distribution corresponding to a multiplicative shift of the standard deviation. The shape and scale parameters of the Weibull distribution must be positive. The two-parameter Weibull is equivalent to a three-parameter Weibull with $\alpha$ = 0. The two-parameter Weibull distribution can arise from fractal particle fragmentation, and for constant particle density a two-parameter Weibull distribution in mass (rather than in particle number) is equivalent to the Rosin distribution \citep{bw95}. The Rosin distribution gives the mass percentage $Y$ of material passed by a sieve of mesh width $d$ as $Y=100\left[1-{\rm exp}\left(-\frac{d^{\beta}}{\gamma}\right)\right]$, where $\beta$ and $\gamma$ are once again the shape and scale parameters of the distribution. In some of the statistical tests described below, the roles of location, shape, and scale parameters are generalized to other distributions, but they retain the same basic significance.

Hypotheses for a chondrule sorting process operating prior to parent body formation include sorting by mass \citep{kmc97}, sorting due to photophoresis \citep{wk06}, sorting due to X-winds \citep{ssl96}, sorting due to disk winds (Teitler, in preparation), and sorting by turbulent concentration \citep[TC;][]{hcd99, hc01, chpd01}. Photophoresis relies on non-uniform desorption of gas molecules from dust grains with non-uniform surface temperatures. Sorting by x-wind or disk winds involves dust grains being entrained in the nascent outflow and then dropping back onto the disk. In the TC scenario, chondrules that encounter turbulence are strongly concentrated in regions of low vorticity where their gas-drag stopping times match the local Kolmogorov eddy turnover time \citep{chpd01}. Possible sources of turbulence include the magnetorotational instability \citep{bh91, bh98}, the Kelvin-Helmholtz instability \citep{go05, c08}, and other mechanisms \citep{cw06, rukr09}.

The relevant sorting parameter for photophoresis sorting is the particle density times the thermal conductivity of the particle \citep{wk06}. The x-wind, disk wind and TC mechanisms involve aerodynamic sorting, which results in sorted distributions that are functions of the product of particle density and particle radius $\rho r$ if particle drag is in the Epstein regime (the limit of large Knudsen number $Kn = \lambda/r \gg 1$, where $\lambda$ is the mean free path of gas molecules), or functions of $\rho r^{2}$ in the Stokes drag regime (for $Kn \ll 1$). An aerodynamic process could produce a distribution in $\rho r^{x}$ for any value of $x$ between 1 and 2, but it is thought that conditions in the early solar nebula correspond to Epstein drag and that aerodynamic sorting should therefore produce a distribution in $\rho r^{x}$ with $x$ very close to 1 \citep{d76}. In general, aerodynamically sorted sets of spherical particles are described by probability density functions $f(z)$ with $z=\rho r^{x}/C$, where $C$ is a normalization factor that is independent of particle properties; it is assumed to be constant for all chondrules in a given chondrite, and to vary from one chondrite to the next. Throughout this paper we take aerodynamic sorting to mean aerodynamic sorting in the Epstein regime ($x = 1$). In the case of TC, the sorting parameter $z$ is the Stokes number $St_{\eta} = \rho r/\rho_{g}c_{s}t_{\eta}$ and the relevant normalization factor $C$ is the local gas density $\rho_{g}$ times the local speed of sound $c_{s}$ times the Kolmogorov eddy turnover time $t_{\eta}$ \citep{chpd01}. Detailed predictions of the sorting distribution function are not available for the mass-sorting, photophoresis, disk wind and x-wind models. Numerical simulations of TC give a prediction for the resulting sorted distribution $f(St_{\eta})$ when a uniform initial distribution of particles is injected into a turbulent region \citep{hc01}. Using radius and density measurements of chondrules disaggregated from several chondrites, we test the hypotheses of sorting by mass, sorting by radius, sorting by density, sorting by a generic aerodynamic process, and sorting by the specific aerodynamic process of TC (under the assumptions that the initial chondrule distribution prior to TC is not too far from uniform over the relevant range of values of $St_{\eta}$, and that parent body processes have not significantly disturbed the chondrule distribution). We also test for sorting as a function of $\rho r^{2}/C$, which would be consistent with aerodynamic sorting in the Stokes regime.

We discuss the chondrule data sets in the next section; in the subsequent sections we describe the test procedures and results, present a summary of our work, and outline some extensions of the current analysis and possibilities for future studies.

\section{CHONDRULE DATA}\label{data}

Measurements of chondrule radius and mass have been made for hundreds of chondrules in various chondrite samples using the freeze-thaw disaggregation technique \citep{pc97, chp99, chpd01}. Chondrule data have also been obtained from other chondrite samples by picking chondrules from available fines. Of these, ten data sets from six different chondrites were selected for testing. Each chondrule has radius measurements along two orthogonal directions. The average of the two radii is then used to calculate the density from the mass, assuming that chondrules are spherical. The aspect ratio of the radii is close to one in all cases. The ten data sets are summarized in Tables \ref{tab:datasets} and \ref{tab:setstats}.

\begin{table}[ht]
 \caption{Chondrule data sets.}
 \label{tab:datasets}
 \begin{tabular}{@{}lcccc}
 \hline
Sample & Chondrite & Chondrite & Chondrules & Method of \\
number & sample & type & weighed / total & chondrule collection \\
\hline
1 & QUE 93030,18 & H3.6 & 100 / 125 & picked \\
2 & ALH 84028,16 & CV3 & 135 / 194 & disaggregated \\
3 & Bjurbšle & L4 & 144 / 150 & picked \\
4 & GRO 95524,6 & H5 & 226 / 300 & disaggregated \\
5 & GRO 95524,7 & H5 & 85 / 86 & picked \\
6 & ALH 85033,22 & L4 & 235 / 300 & disaggregated \\
7 & ALH 85033,23 & L4 & 54 / 59 & picked \\
8 & ALH 85033,24 & L4 & 39 / 54 & picked \\
9 & Allende,1 & CV3 & 282 / 287 & disaggregated \\
10 & Allende,2 & CV3 & 29 / 126 & disaggregated \\
 \hline
 \end{tabular}
\end{table}

\begin{table}[ht]
 \caption{Chondrule data set statistics.}
 \label{tab:setstats}
 \begin{tabular}{@{}lcccccccc}
 \hline
Sample & Range of & Mean & Median & Stdev & Range of & Mean $\rho r$ & Median $\rho r$ & Stdev $\rho r$ \\
number & $r$ (cm) & $r$ (cm) & $r$ (cm)  & $r$ (cm) & $\rho r$ (g cm$^{-2}$) & (g cm$^{-2}$) & (g cm$^{-2}$)  & (g cm$^{-2}$) \\
\hline
1 & 0.0120-0.116 & 0.0321 & 0.0282 & 0.0161 & 0.0200-0.350 & 0.0870 & 0.0777 & 0.0481 \\
2 & 0.0143-0.183 & 0.0466 & 0.0394 & 0.0244 & 0.0174-0.411 & 0.108 & 0.105 & 0.0570 \\
3 & 0.0148-0.141 & 0.0538 & 0.0513 & 0.0270 & 0.0170-0.507 & 0.164 & 0.156 & 0.0934 \\
4 & 0.00750-0.0663 & 0.0257 & 0.0235 & 0.0110 & 0.00558-0.226 & 0.0747 & 0.0697 & 0.0353 \\
5 & 0.0164-0.0994 & 0.0464 & 0.0456 & 0.0175 & 0.0146-0.341 & 0.105 & 0.0996 & 0.0504 \\
6 & 0.00870-0.0949 & 0.0231 & 0.0192 & 0.0130 & 0.00952-0.190 & 0.0696 & 0.0659 & 0.0288 \\
7 & 0.0156-0.0714 & 0.0330 & 0.0297 & 0.0139 & 0.0210-0.240 & 0.102 & 0.0930 & 0.0376 \\
8 & 0.0213-0.0653 & 0.0356 & 0.0331 & 0.0121 & 0.0184-0.225 & 0.0972 & 0.0956 & 0.0457 \\
9 & 0.0133-0.455 & 0.0456 & 0.0390 & 0.0322 & 0.00994-0.343 & 0.0897 & 0.0798 & 0.0529 \\
10 & 0.0137-0.198 & 0.0459 & 0.0316 & 0.0372 & 0.0453-0.518 & 0.143 & 0.140 & 0.0998 \\
 \hline
 \end{tabular}
\end{table}

There are several potential sources of error in the data sets. There is always a possibility that chondrules were missed or destroyed during the disaggregation process. There was no attempt made to collect clastic fragments \citep{h78a}. It would be expected that the very smallest chondrules would be most likely to be missed, while destruction of chondrules could affect all sizes. We suspect that the disaggregation process separated any compound chondrules initially present, as no compound chondrules were identified in the samples. We expect that the fraction of compound chondrules was small enough in each sample that their separation does not affect our results. The largest error in the measurements occurs because the 3-dimensional chondrule is only measured in two dimensions. There are very few chondrules that vary significantly from round. We did our best to minimize this potential source of error by measuring the largest and smallest dimension, so the average is closer to the average of all three dimensions. Errors may also be introduced during the weighing. All chondrules were weighed twice and, if the measurements were not in agreement, the chondrule was weighed a third time. Samples were generally re-weighed if the first two weighings disagreed by more than 0.0003 g, but in many cases a third weighing was done even for smaller disagreements in the first two weighings. All measurements were taken without having the previous measurement at hand. In general, it is likely that errors due to the accuracy of measurements will cancel out over the large number of samples. Any systematic errors that affect the whole population of chondrules will not affect the interpretation of the data for this study.

The Allende,2 sample (set 10) only had 29 chondrules weighed out of 126 disaggregated, leaving open the possibility of bias in the selection of chondrules that were weighed. The other disaggregated data sets have at least 70\% of the chondrules weighed, diminishing the effects of possible bias.

Data sets that were picked may suffer from another form of selection bias. We expect that picked sets of chondrules will preferentially sample the large-size end of the disaggregated distribution, as the larger chondrules are more easily picked from available fines. The presence of pairs of disaggregated and picked samples from the same chondrite offers the opportunity to test for this selection bias. Support for the hypothesis of picking bias can be found in Table 2, which shows that the typical values of $r$ and $\rho r$ of the picked data sets are consistently higher than those of the disaggregated data sets in cases where both types of data sets are available from a single chondrite. Before proceeding with our analysis we investigate possible biases in the picked data sets. 

\subsection{Comparisons of picked and disaggregated data sets}\label{comp}

The two-sample Anderson-Darling test \citep{p76} was employed to investigate possible picking biases by comparing the distributions of picked and disaggregated chondrules in sets 4 and 5; the two-sample Anderson-Darling test was also employed in comparing the two disaggregated data sets 9 and 10 to test for possible weighing selection bias in set 10. The $k$-sample Anderson-Darling test \citep{ss87} was used for the three data sets from ALH 85033. The Anderson-Darling statistic is a goodness-of-fit test that is sensitive to all differences in distribution, with better sensitivity to differences in the distribution tails and better power performance (lower probability of false positives indicating consistency of data sets that are actually drawn from different distributions) in many situations than its competitors \citep{s74}. The test was performed for the sorting parameters $r$, $\rho$, $\rho r$, $\rho r^{2}$, and $\rho r^{3}$.

\begin{figure}[ht] \label{fig:1}
	\centering
		\includegraphics[width=5in]{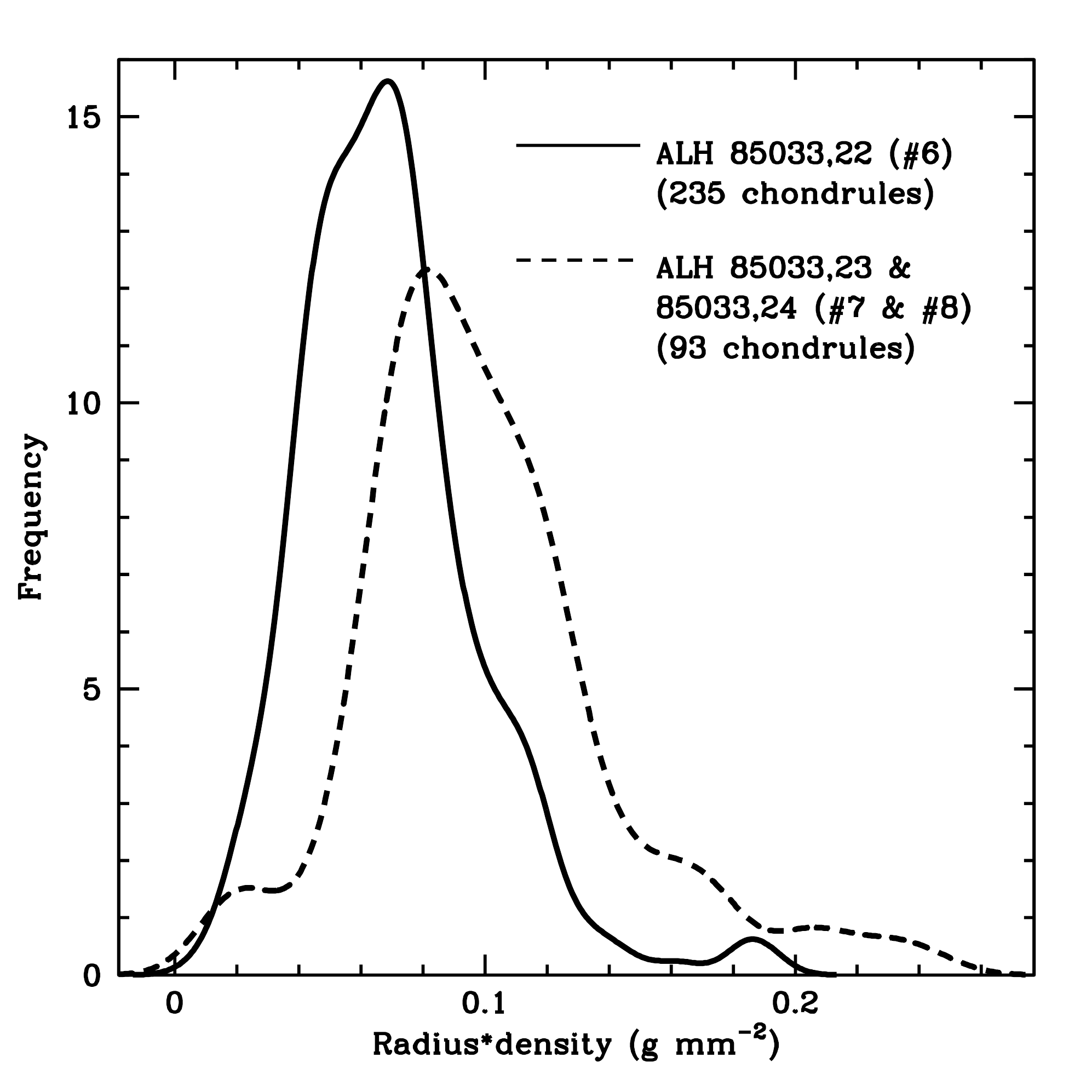}
	\centering
\caption{Comparison of kernel density estimates of chondrule data sets from ALH 85033.}
\end{figure}

The Anderson-Darling tests for consistency of picked and disaggregated sets of chondrules from a given chondrite gave rejections for data sets 4 and 5 (GRO 95524) and sets 6-8 (ALH 85033). Further testing yielded a two-sample acceptance of sets 7 and 8 for all sorting parameters. We generated Gaussian kernel density estimates (generalizations of histograms that assign a Gaussian function to each data point and sum to give estimates of parent distribution curves) of the $\rho r$-distributions of data set 6 and the combination of sets 7 and 8, which are compared in Fig. \ref{fig:1}. The bandwidth of the Gaussian used in each case was selected using Silverman's rule of thumb \citep{s86} and kernel density estimates were generated using the statistical package R \citep{r07}. Note that the data sets are not normalized, since all three come from the same chondrite and are thus assumed to share a common normalization factor. It is clear from this comparison that the sets are not consistent in their $\rho r$-distributions, showing differences even in the mean values, as expected from the data in Table 2. Similar differences appear in comparisons of kernel density estimates of the distributions of the sets as functions of other combinations of $\rho$ and $r$.

As a result of this analysis, we retained the disaggregated data sets 4 and 6 for use in further tests as representative of GRO 95524 and ALH 85033, discarding the picked data sets 5, 7 and 8. We also discarded the picked data sets 1 and 3 on the grounds that the picked data sets had shown a clear and consistent bias in the cases we had been able to check. Finally, the two-sample Anderson-Darling test of the disaggregated sets 9 and 10 indicated inconsistency between sets 9 and 10, so we also discarded set 10 on grounds of having too low a weighed fraction (approximately 23\%). We hypothesize that there was some bias in the selection of chondrules that were weighed and that weighing more of the chondrules in set 10 would remove the inconsistency of sets 9 and 10; for the present, we proceed with sets 2, 4, 6 and 9.

\section{TEST PROCEDURES}\label{test}

\subsection{Graphical methods of comparison}

\begin{figure}
\centering
\begin{tabular}{cc}
\epsfig{file=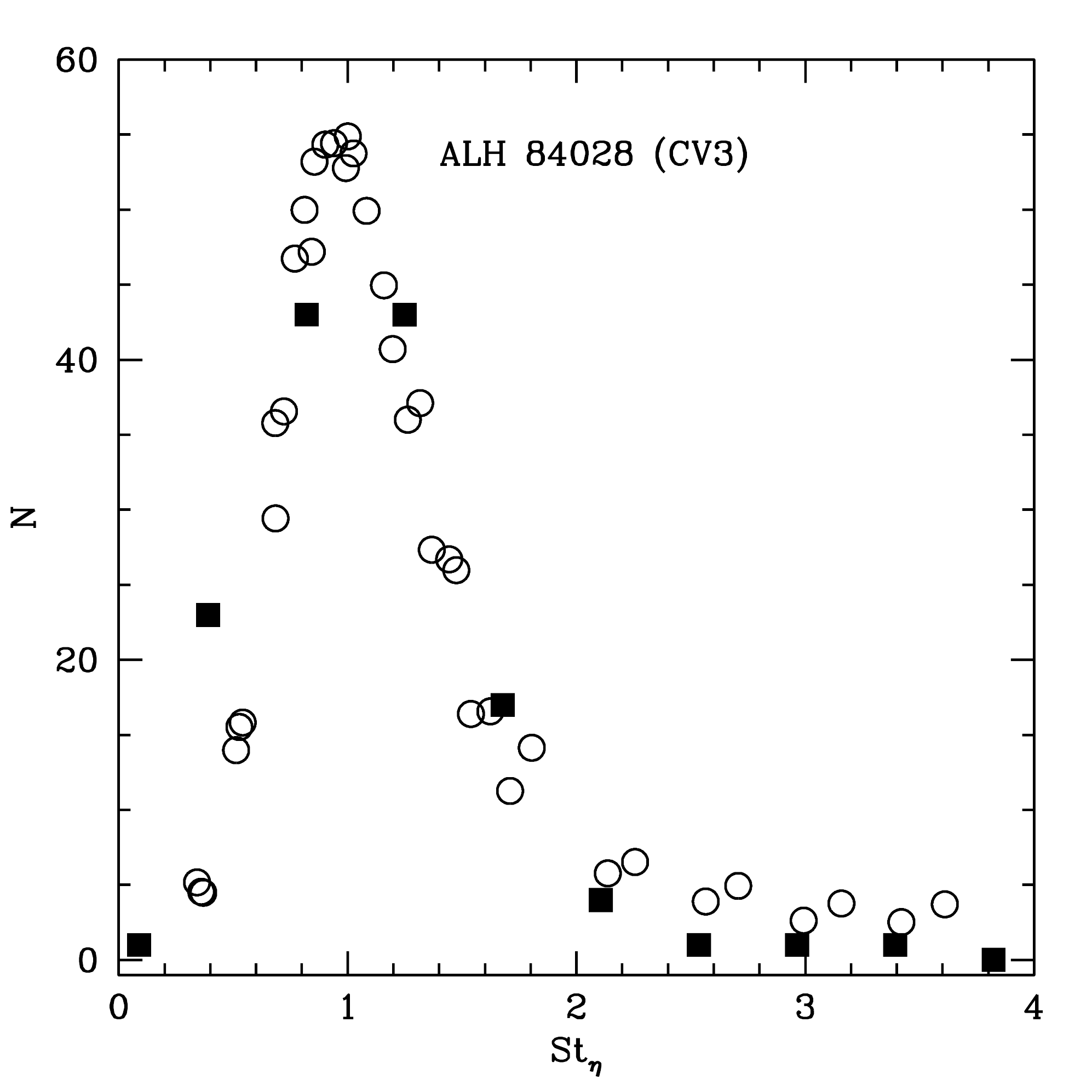,width=0.5\linewidth,clip=} &
\epsfig{file=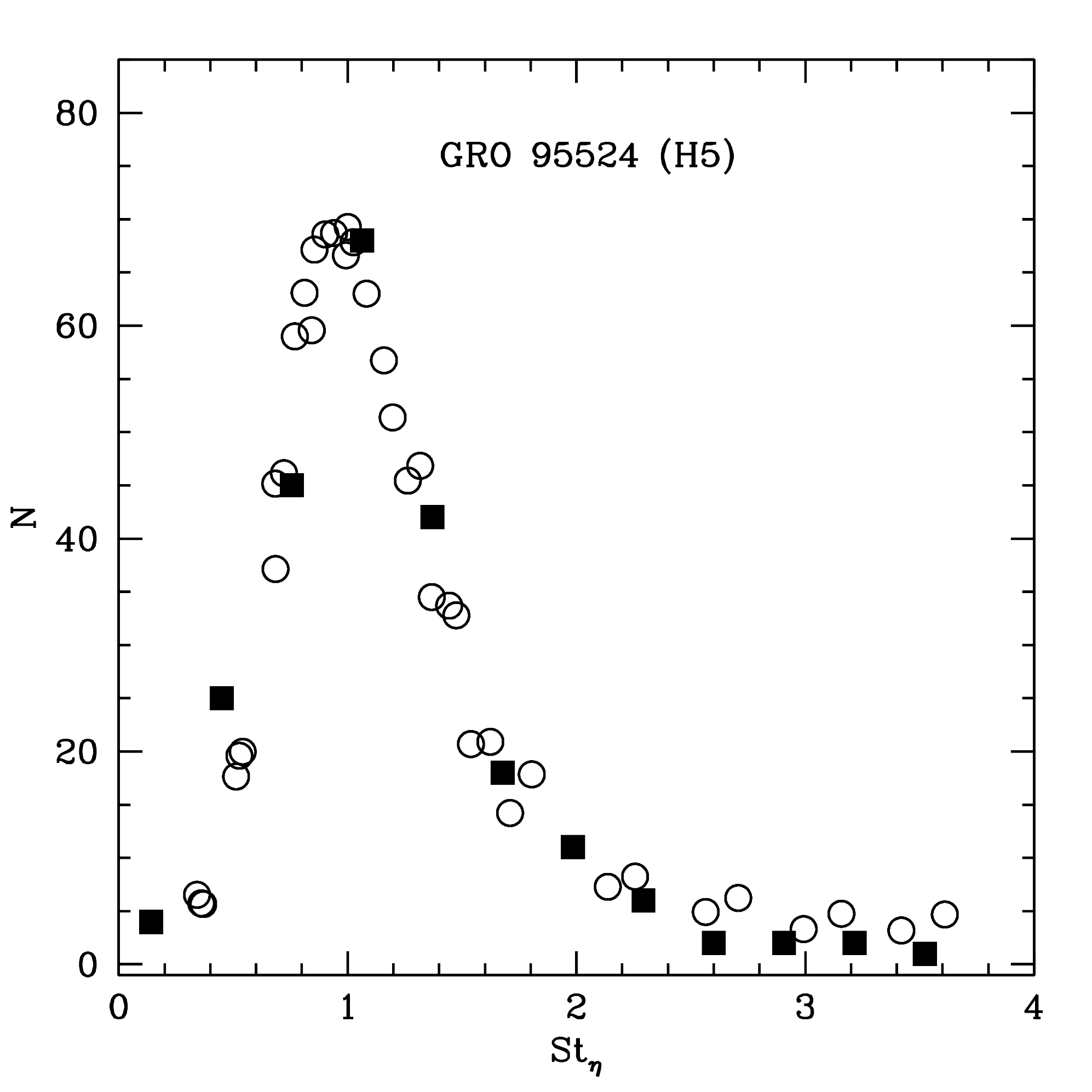,width=0.5\linewidth,clip=} \\
\epsfig{file=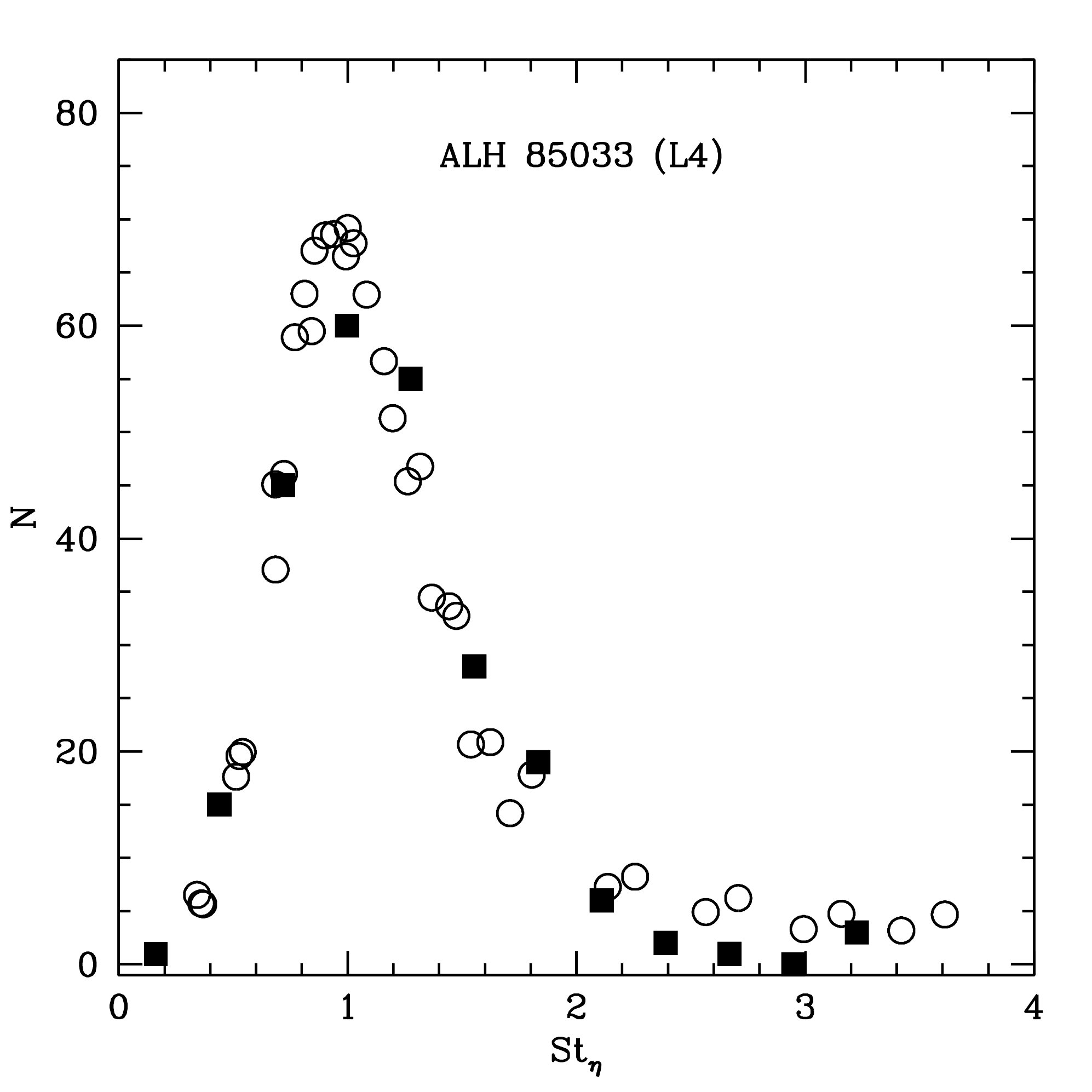,width=0.5\linewidth,clip=} &
\epsfig{file=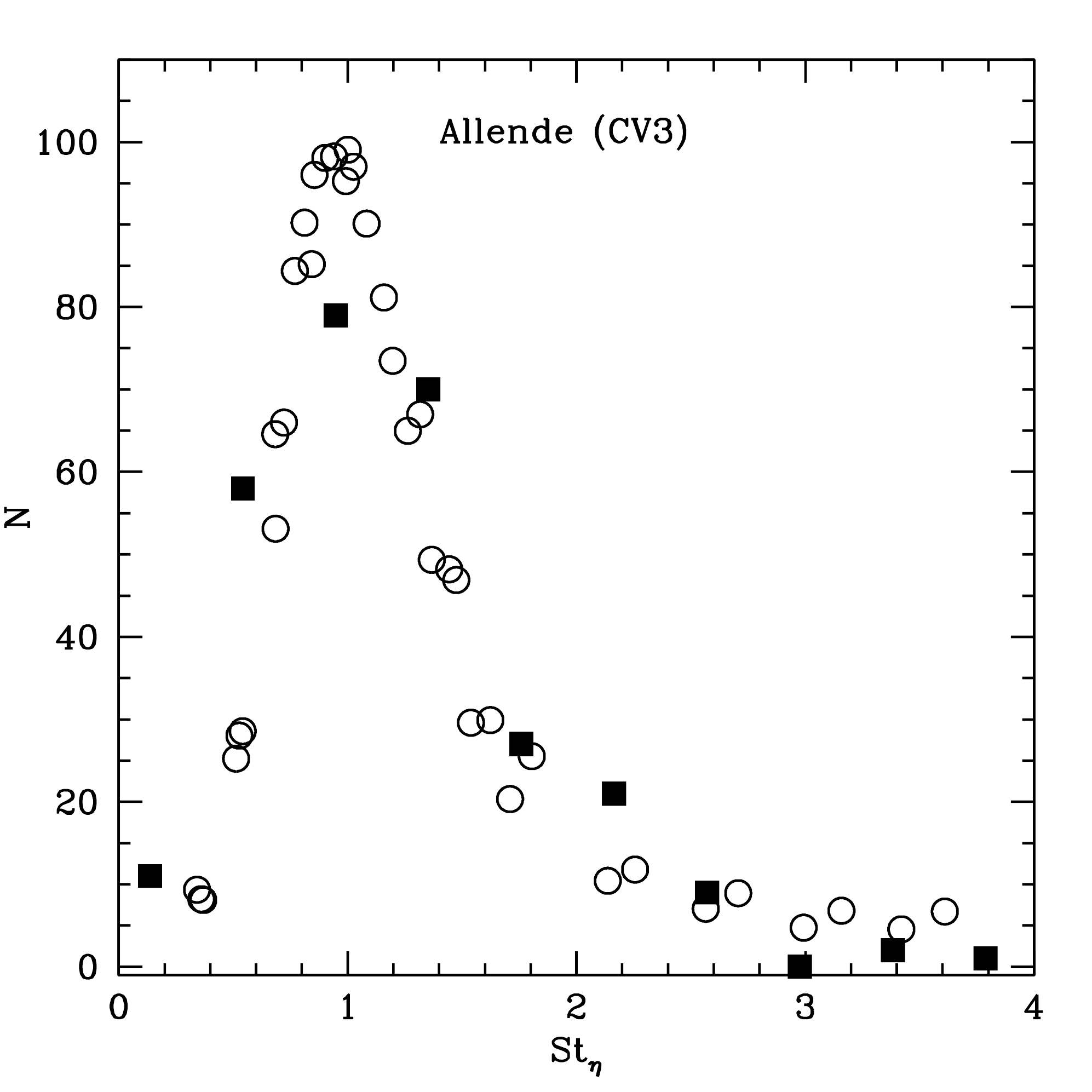,width=0.5\linewidth,clip=} \\
\end{tabular}
\caption{Comparison of size-density distributions of four sets of disaggregated chondrules (solid symbols) and predictions of turbulent concentration simulations (open symbols).  The panels for GRO 95524 and ALH 85033 are essentially the same as in Figure 2 of Cuzzi et al. (2001).}
\end{figure}

The concentrations predicted by numerical simulations of TC \citep{hc01} at a discrete set of values of the Stokes number are compared with histograms of the $\rho r$ distributions for data sets 2, 4, 6 and 9 in Fig. 2. This figure is very similar to Fig. 2 of \citep{chpd01}, but here we add plots for data sets from the chondrites ALH 84028 and Allende, and we do not include plots for the chondrites OTTA 80301 or Bjurb\"{o}le since both have only picked chondrule data sets available.

\begin{figure} \label{fig:3}
\centering
\begin{tabular}{cc}
\epsfig{file=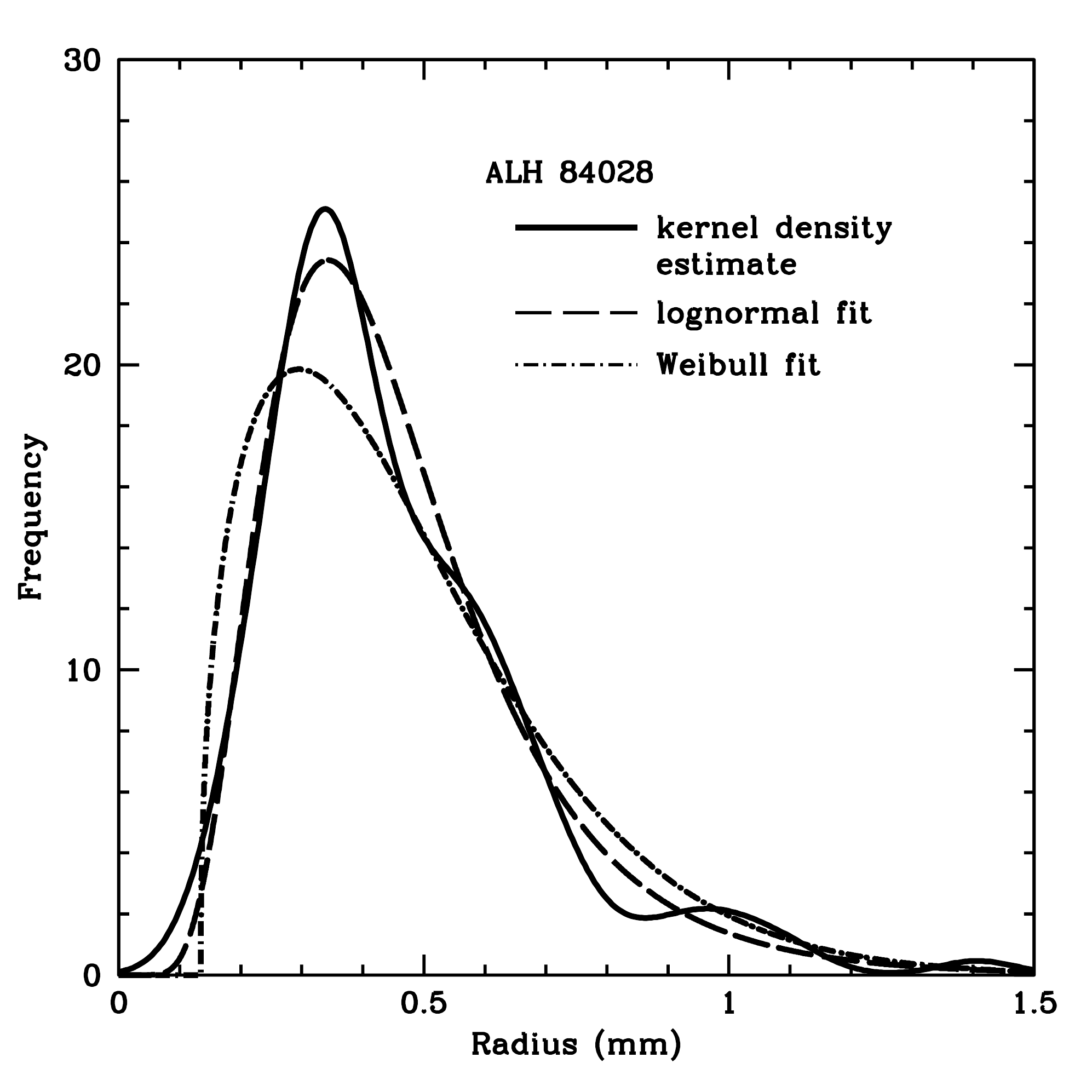,width=0.5\linewidth,clip=} &
\epsfig{file=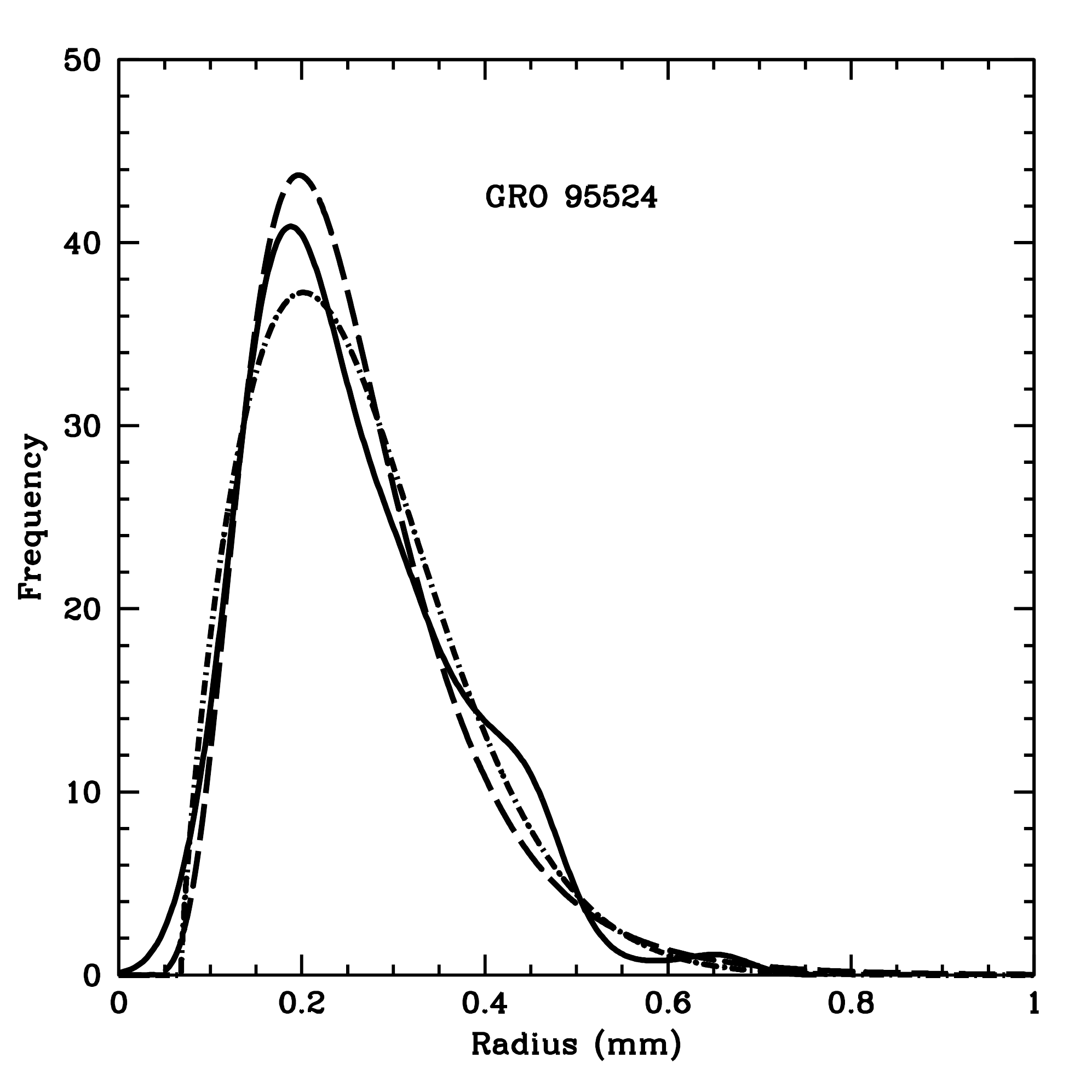,width=0.5\linewidth,clip=} \\
\epsfig{file=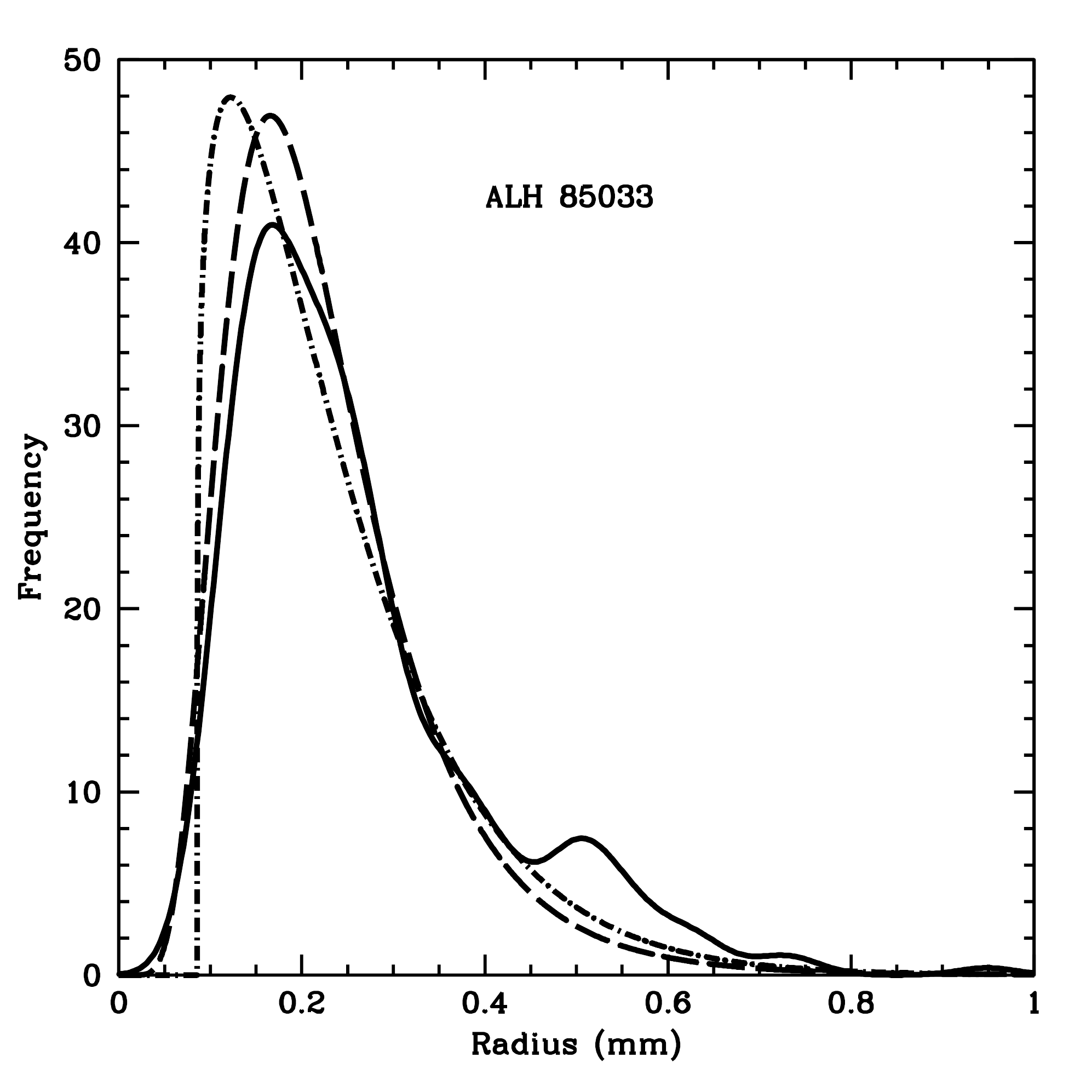,width=0.5\linewidth,clip=} &
\epsfig{file=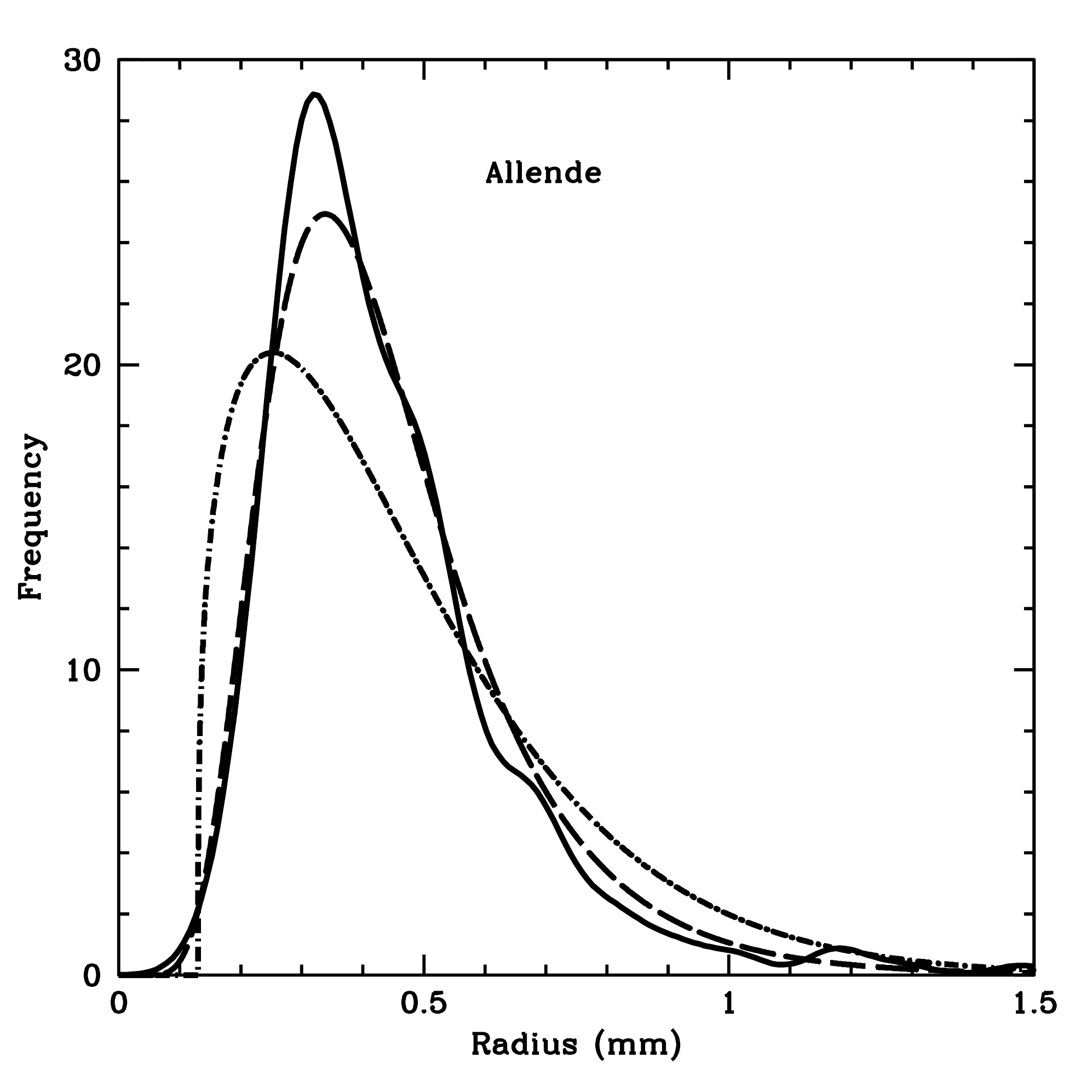,width=0.5\linewidth,clip=} \\
\end{tabular}
\caption{Comparison of kernel density estimates of size ($r$) distributions of four chondrule data sets, fitted lognormal and 3-parameter Weibull distributions.}
\end{figure}

Kernel density estimates of the size distributions of data sets 2, 4, 6 and 9 are compared with lognormal and three-parameter Weibull fits in Fig. 3. The comparison suggests that the lognormal and Weibull fits are not simultaneously satisfactory for all chondrule data sets, even if individual chondrule data sets seem to conform to a lognormal or Weibull fit. This conclusion is supported by the quantitative tests described below.

\subsection{Multiple comparisons procedure}
All further goodness-of-fit tests in this investigation were conducted at the 95\% significance level using the Benjamini-Hochberg procedure for multiple comparisons \citep{bh95}. Multiple comparisons procedures involve running several two-sample tests on pairs of samples from a pool of data sets, then adjusting the results in some manner so that the significance level of all test results considered simultaneously is very close to the significance level used in the initial two-sample tests. In the case of the Benjamini-Hochberg method, the results of the two-sample tests are adjusted as follows: let $P_{1} < P_{2} < \dots < P_{n}$ denote the ordered $p$-values of the $n$ two-sample tests (the $p$-value of a test is the smallest level at which the null hypothesis is rejected, or one minus the maximum significance level at which the null hypothesis is accepted; a $p$-value of 0.05 corresponds to a significance level of 95\%). For a desired significance level 1-$\alpha$, let $k$ denote the largest $i$ such that $P_{i} \leq i \alpha/n$; then the tests corresponding to the $p$-values $P_{1} < P_{2} < \dots < P_{k}$ return rejections of the null hypothesis (that is, the two samples are inconsistent). The Benjamini-Hochberg method was selected because it does a better job of preserving test power than alternatives like the Bonferroni method \citep{wjt99}, which takes a more conservative approach that gives a significance level at least as high as that of the individual two-sample tests but typically results in low power.

\subsection{Tests for goodness-of-fit to simple distributions}
Chondrule data from four different chondrites were tested for consistency with being drawn from lognormal parent distributions using the Anderson-Darling statistic. The Anderson-Darling statistic was selected for its ability to test whether a sample is drawn from a lognormal distribution without specifying the mean or variance in advance (i.e. its ability to test a composite hypothesis). In this case, the composite hypothesis was that each chondrite sample was drawn from a lognormal distribution $f(z)$ with mean $\mu$ and variance $\sigma$ matched to the sample means and variances of each chondrite sample. The simpler Kolmogorov-Smirnov (K-S) test \citep{hw73}, like the Anderson-Darling test, is sensitive to all differences in distribution, but it cannot immediately tackle a composite null hypothesis. That is, setting parameters of the reference distribution using sample values affects the critical region of the K-S test, changing the correspondence between different values of the K-S statistic and significance levels. In the present study, the Anderson-Darling tests of each chondrule data set were run for lognormal number density distributions in $r$, $\rho$, $\rho r$, $\rho r^{2}$, and $\rho r^{3}$ for completeness.

The chondrule data sets were also tested for consistency with a best-fit three-parameter Weibull distribution in $r$. Parameter estimation by the maximum-likelihood method can fail in cases where one of the parameters describes a bound on possible data values; since the location parameter $\alpha$ of the three-parameter Weibull distribution gives a lower bound on the range of possible data values, we use the maximum product of spacings method \citep{ca83} instead of the maximum-likelihood method for parameter estimation. The maximum product of spacings method has good accuracy against a wide variety of alternatives when the shape parameter $\beta$ is larger than 1 \citep{c09}, which is the case for all the data sets. After estimating the parameters for each data set, the goodness-of-fit was investigated using the Anderson-Darling test for the three-parameter Weibull distribution \citep{ls94}.

\subsection{Tests for sorting by various single parameters and by turbulent concentration}
Statistical tests of the general hypothesis of aerodynamic sorting and the specific hypothesis of sorting by TC are complicated by the unknown normalization factor $C$ that relates the sorting parameter ($St_{\eta}$ for the TC hypothesis) to the measured chondrule radius-density product $\rho r$. This normalization factor is based on local gas conditions that vary from one chondrite sample to the next, but is assumed to be constant for all chondrules from a given chondrite sample. Testing is further complicated by the fact that the tests employed must be distribution-free, since a general hypothesis of aerodynamic sorting does not specify any particular form for the parent distribution function $f(z = \rho r/C)$ and the specific hypothesis of sorting by TC specifies a form for the parent distribution function that doesn't correspond to any simple analytic curve. Similar difficulties hold for testing a general hypothesis of sorting by mass or any other single sorting parameter (again possibly normalized by a factor $C$ that varies from chondrite to chondrite).

The K-S test is distribution-free, but can only be used with simple null hypotheses in which the reference distribution is fully specified; the need to account for an unknown normalization factor means that we cannot use the K-S test. Conversely, the Anderson-Darling test can handle parameter estimation from data sets, but is not distribution-free, and the $k$-sample Anderson-Darling test on the chondrule data sets after shifting the sample means to align loses so much power as to render it useless: All sorting parameters are accepted, with the exception of sorting by $\rho$, which does produce a rejection. A suitable alternative is the Miller jackknife test \citep{q49, m68}, which has the disadvantage that it is not sensitive to general differences in distribution, unlike the Anderson-Darling test: for instance, the Miller test is insensitive to differences in the parent distribution location parameters. The Miller jackknife is a test for differences in the parent distribution scale parameters of two sets of data which is asymptotically distribution-free, meaning that the result is independent of the specific form of the parent distribution for large data sets. The null hypothesis is that two sets of data have the same form of the parent distribution function with the same scale parameter, but possibly with different location parameters. The alternative hypothesis is that two sets of data have the same form of the parent distribution function with different scale parameters and possibly different location parameters. The Miller test can give a false positive (non-rejection of the null hypothesis) for two sets of data drawn from completely different parent distributions, possibly with different scale factors. The Miller test was employed both to test the hypotheses of sorting by various parameters and to test the hypothesis of sorting by TC.

The procedure used in this paper involved running the Miller test on the logarithm of the chondrule properties $z$ = $r$, $\rho$, $\rho r$, $\rho r^{2}$, and $\rho r^{3}$. For instance, to test whether the chondrule data sets are all sorted according to some distribution $f(z = r/C)$ we take the sets of $r$-data, take the logarithm of every data point, run the Miller test on the ${\rm log}(r)$-data for pairs of chondrule data sets, then combine the results using the Benjamini-Hochberg procedure. Sorting according to some distribution $f(z = r/C)$ corresponds to sorting according to some distribution $g\left({\rm log}(z)\right) = g\left({\rm log}(r)-{\rm log}(c)\right)$; differences in the normalization factor $C$ from one data set to the next are translated into shifts in the location parameters, to which the test is insensitive. Thus any rejection produced by the Miller test procedure used here signals that the chondrule sets being compared do not share a common parent distribution (up to differences in the normalization factor) in the sorting parameter $z$ = $r$, $\rho$, $\rho r$, $\rho r^{2}$, or $\rho r^{3}$ being tested.

The Miller procedure was also used to test chondrule $\rho r$-data sets for compatibility with the parent distribution predicted by the TC hypothesis. The test used a set of 1,000,000 data points drawn from the TC parent distribution, and once again was run on the logarithms of the data sets so that differences in normalization factors would not produce spurious rejections. A smooth interpolation of the $Re$ = 80 data points from earlier TC simulations was used as an approximation to the TC parent distribution; note that the shape of the parent distribution is insensitive to the Reynolds number \citep{hc01}.

Finally, as a check on the sensitivity of the Miller procedure, we tested chondrule data sets 2, 4, 6 and 9 for consistency with simulated data sets of 1,000 data points each drawn from lognormal and three-parameter Weibull distributions fitted to the individual chondrule data sets.
 
\subsection{Monte Carlo investigation of the Miller test}
The effectiveness of the Miller procedure for small data sets was investigated using a series of Monte Carlo tests for significance and for power against various alternative hypotheses. The tests pitted 1000 sets of 50 data points each drawn from the TC parent distribution against 1000 other sets of 50 data points each drawn from the TC parent distribution multiplied by a scaling factor $\Delta$ (a scaling of the horizontal axis or $St_{\eta}$ values in the interpolated TC prediction curve shown in Fig. 1), for a few different values of $\Delta$. This has the effect of changing the variance of the second sample's parent distribution by a factor $\Delta^{2}$. When $\Delta^{2} = 1$, both samples are drawn from the same parent distribution, so the null hypothesis is true. In this case the significance level of the test, which is the probability of a correct acceptance given that the null hypothesis is true, is given by the percentage of non-rejections. The power of a test is similarly defined as the probability of a correct rejection given that the null hypothesis is false---in this case, given that the two samples are drawn from different parent distributions. The power depends on the parent distributions of the two samples being tested and the significance level at which the test is conducted. In cases where $\Delta^{2} \neq 1$ our null hypothesis is false and the observed power for the Miller test conducted at the 95\% significance level is given by the percentage of rejections. These tests were not conducted using the Benjamini-Hochberg procedure, since the goal was to investigate the power and significance performance of the Miller test on finite data sets.

\section{RESULTS}\label{res}

The Anderson-Darling tests for lognormal parent distributions in $r$, $\rho$, $\rho r$, $\rho r^{2}$, and $\rho r^{3}$ were conducted on the data sets from the four chondrites. The tests were run twice, once with set 6 included in the collection of data sets, and once using only sets 2, 4 and 9. The hypothesis of consistency with a lognormal parent distribution had multiple rejections in all cases except for the lognormal in $\rho r^{2}$. The Anderson-Darling tests for three-parameter Weibull parent distributions in $r$ for the chondrule data sets also gave multiple rejections of the group hypothesis in both sets of tests.

A previous investigation with a somewhat different collection of 7 chondrule data sets from 6 chondrites \citep{tpch09} gave the following results: the test for sorting by $\rho r$ had 0 rejections; the test for sorting by $r$ had 3 rejections, all involving the Bjurb\"{o}le data set; the test for sorting by $\rho$ had 4 rejections, not all involving any one sample; and the tests for sorting by $\rho r^{2}$ and $\rho r^{3}$ each had 6 rejections, all involving the Bjurb\"{o}le data set. The differences in the collection of data sets considered in the previous investigation are as follows: the previous investigation did not include data sets 6, 8 and 10 of the current collection, the previous investigation did include a data set of 100 chondrules picked from OTTA 80301 not included in the current collection, and the previous investigation retained the picked data sets 1, 3, 5 and 7 throughout the analysis. The previous investigation concluded that the OTTA 80301 data set was problematic, most likely due to a much smaller fraction of chondrules being measured from that sample than in other cases (see \citealt{tpch09} for details), so that data set was excluded from the current investigation.

\begin{table}[ht]
 \caption{Miller test of sorting parameter.}
 \label{tab:mill}
 \begin{tabular}{@{}lccc}
 \hline
Sorting & Current rejections & Current rejections & Previous \\
parameter & (with data set 6) & (without data set 6) & rejections$^{a}$ \\
\hline
$r$ & 0 & 0 & 3 \\
$\rho$ & 1 & 1 & 4 \\
$\rho r$ & 1 & 0 & 0 \\
$\rho r^{2}$ & 0 & 0 & 6 \\
$\rho r^{3}$ & 0 & 0 & 6 \\
\hline
\end{tabular}
\end{table}
\noindent $^{a}${\footnotesize from Teitler et al. (2009)}

The current round of Miller tests of sorting parameter ran on paired chondrule sets, giving 6 comparisons when set 6 was included and 3 comparisons when it was not. The tests for sorting by $r$ had 0 rejections for both collections; the test for sorting by $\rho$ had 1 rejection for both collections; the test for sorting by $\rho r$ had 1 rejection involving the ALH 85033,22 data set (set 6) when it was included and 0 when it was not; the test for sorting by $\rho r^{2}$ had 0 rejections; and the test for sorting by $\rho r^{3}$ had 0 rejections. The results of the current and previous rounds of testing are summarized in Table 3.

\begin{table}[ht]
 \caption{Monte Carlo test of Miller's jackknife.}
 \label{tab:mcmill}
 \begin{tabular}{@{}lccc}
 \hline
$\Delta^{2}$ & Rejection percentage \\
\hline
1 & 5.9\%$^{a}$ \\
2 & 73.2\%$^{b}$ \\
4 & 99.7\%$^{b}$ \\
\hline
\end{tabular}
\end{table}
\noindent $^{a}${\footnotesize The observed significance level is 1 minus the probability of a false negative, corresponding to 1 minus the rejection percentage for the case $\Delta^{2} = 1$ (both samples drawn from the same parent distribution).}

\noindent $^{b}${\footnotesize The observed power level is the probability of a true negative, corresponding to the rejection percentage for any case where $\Delta^{2} \neq 1$ (samples not drawn from the same parent distribution). The power depends on the dissimilarity of the samples' parent distributions.}

The Miller tests of chondrule sets 2, 4, 6 and 9 compared with a simulated set of 1,000,000 data points drawn by Monte Carlo techniques from the TC parent distribution in $\rho r$ gave a rejection due to set 6. Upon removing set 6 from the collection of data sets, the Miller tests of the chondrule sets against the simulated TC data set gave no rejections. The Miller tests of the chondrule data sets against simulated lognormal and Weibull data sets also gave no rejections.

The results of the Monte Carlo tests for the significance and power of the Miller procedure for testing the TC hypothesis are summarized in Table 4. The percentage of rejections (out of 1000 pairings) is reported for each alternative hypothesis, represented by a data set drawn from the TC parent distribution multiplied by a scaling factor $\Delta$. When $\Delta^{2} = 1$, the paired data sets are drawn from the same parent distribution, so the null hypothesis is true and any rejection of the null hypothesis is a false rejection. The effective significance level (1 minus the percentage of false rejections) of 94.1\% is very close to the target value of 95\%, suggesting that the procedure works quite well for sets of 50 data points. When $\Delta^{2} \neq 1$ the paired data sets are not drawn from the same parent distribution, so the null hypothesis is false and any positive (acceptance of the null hypothesis) is a false positive. The results of the Monte Carlo tests with $\Delta^{2} \neq 1$ highlight the good power performance (1 minus the percentage of false positives) of the Miller test. We conclude that the fact that the Miller test is only asymptotically distribution-free is not a concern for chondrule data sets of the sizes considered in our study.

\section{DISCUSSION}\label{disc}

The Anderson-Darling test was employed to demonstrate that the chondrule data as a group are inconsistent with lognormal distributions in various sorting parameters, and are inconsistent with three-parameter Weibull distributions in size. The data sets as a group are consistent with a lognormal distribution in $\rho r^{2}$. We suspect that this result would change with a larger collection of chondrule data sets. These results are independent of the inclusion or exclusion of data set 6. These results do not rule out the hypothesis of a Rosin distribution.

The Miller tests comparing chondrule data sets and simulated lognormal and Weibull data sets gave no rejections. This is unsurprising, since the lognormal and Weibull distributions were fitted to the individual chondrule data sets, including fits of the scale parameters of the distributions. These results confirm that the Anderson-Darling test is sensitive to differences in distributions that are undetectable with the Miller test. Evidently the Anderson-Darling test is preferable in cases where it can be applied.

The previous series of Miller tests of sorting parameters ($r$, $\rho$, $\rho r$, $\rho r^{2}$, and $\rho r^{3}$) reported in \citet{tpch09} showed a clear preference for sorting by $\rho r$, but made use of several picked chondrule data sets that we do not include here since the picked sets seem biased. After removing these problematic data sets, the results of the Miller test of sorting parameters ($r$, $\rho$, $\rho r$, $\rho r^{2}$, and $\rho r^{3}$) are inconclusive. Data sets 2, 4 and 9 are consistent as a group with sorting by $\rho r$, but including data set 6 results in a rejection of the hypothesis of sorting by $\rho r$ due to the pairing of sets 6 and 9. Similarly, the hypothesis of sorting by $\rho$ gives a rejection due to the pairing of sets 2 and 4. All four data sets are consistent with sorting by $r$, $\rho r^{2}$, and $\rho r^{3}$. The Miller test also showed that data set 6 is inconsistent with the TC hypothesis, but that sets 2, 4 and 9 are consistent as a group with being drawn from the predicted TC curve. Data set 6 is evidently dissimilar from the data sets from other chondrites in its distribution as a function of $\rho r$ and its rejection of the TC hypothesis. It remains unclear whether data set 6 is representative of ALH 85033, in which case the TC hypothesis (and sorting by $\rho r$ in general) does not appear to fit the data from ALH 85033, or if some form of error has affected data set 6.

The sorting parameter analysis would benefit from including more data sets from additional chondrites. Our previous analysis of 7 chondrule data sets \citep{tpch09} indicated a clear preference for sorting by $\rho r$; this specific result cannot be trusted since it is based in part on data sets that we have since concluded are systematically biased, but the result suggests that a clear preference for some sorting parameter may appear if the present analysis is extended beyond the four chondrule data sets considered in this study. Assuming that chondrules and other chondrite components experienced a common sorting process, analysis of other, non-chondrule components of chondrites would also be helpful in distinguishing between various sorting parameters, since the density contrast between chondrules and other components (e.g. metal chondrules, metallic grains, and sulfide grains) is considerably larger than the density contrast between different chondrules \citep{d76, sh93, sl93, kmc97, bas98, as99, kmch99}.

\section{CONCLUSIONS}\label{concs}

Our investigation shows that the chondrule distributions from the chondrites we studied are not well-fit by lognormal or three-parameter Weibull functions, except for a lognormal in $\rho r^{2}$, which does appear to be consistent with the four chondrule data sets considered here. A particular chondrule data set may appear similar to a lognormal or Weibull function when examined by eye, and may even appear consistent with a lognormal or Weibull function when examined using an Anderson-Darling test, but in all cases other than lognormal in $\rho r^{2}$ our tests found multiple rejections---in other words, there are chondrule data sets for which lognormal and Weibull functions are statistically not good fits to the chondrule distributions. Note that \citet{tpch09} found that a somewhat different collection of chondrule data sets was inconsistent with a lognormal distribution in $\rho r^{2}$, but the collection analyzed there included several picked data sets that our current analysis suggests are biased. It is unclear whether consistency with a lognormal distribution in $\rho r^{2}$ is a real feature of chondrule data sets or if that hypothesis will also be rejected with further analysis of more chondrule data sets.

The hypotheses of various common sorting parameters ($r$, $\rho$, $\rho r$, $\rho r^{2}$, and $\rho r^{3}$) remain open and unconstrained, and would probably also benefit from further analysis of more chondrite data sets, particularly sets of non-chondrule components.

Sorting by TC should produce chondrule data sets that are not rejected in Miller tests comparing chondrule data to simulated TC data and in Miller test of pairs of chondrule data as functions of $\rho r$. We obtain consistency with sorting by $\rho r$ (and all other sorting parameters considered, except for $\rho$) and consistency with sorting by TC for the collection of data sets 2, 4 and 9, but not for the collection of data sets 2, 4, 6 and 9. We cannot at this time explain the apparent discrepancies between sets 2, 4 and 9, and set 6. The specific hypothesis of sorting by TC is consistent with the bulk of the chondrule data, so the TC hypothesis remains open, while the specific hypotheses of lognormal or Weibull parent distributions appear less likely based on the results of the Anderson-Darling tests. Without predictions of sorting distributions we cannot run similar tests of other specific sorting mechanisms.

The test procedures described in this paper can be used to determine whether various chondrite components (e.g. chondrules, CAIs and metal and sulfide grains) are consistent with sorting by various mechanisms, given radius and density data for the relevant particles. In particular, the Miller procedure can be used to test for sorting by various parameters (r, $\rho r$, etc.). This approach cannot distinguish between different aerodynamic sorting mechanisms; such a distinction is only possible if predicted distribution functions for each mechanism can be found and compared to chondrule data. In principle, given thermal conductivity data for the chondrules, the Miller procedure could also be used to test the hypothesis of sorting by photophoresis. For sets of data on multiple components from a single chondrite, the more powerful Anderson-Darling test can be used to test the assumption of a common sorting process, since the normalization factor would not differ between components from a given chondrite. Various components can also be compared across chondrites using the Miller test. These test procedures appear to be well-suited to analysis of sets of a few dozen data points or more, and have the flexibility needed to investigate various hypotheses concerning chondrule sorting processes.

We plan to extend the current analysis by incorporating additional chondrule data sets, as well as data from CAIs, metal and sulfide grains, and other chondrite components. The methods described in this paper can also be used to characterize other sets of particles, such as terrestrial spherules from giant impacts or lunar regolith spherules. In particular, these techniques could be used to compare chondrules and lunar crystalline spherules, which have been proposed as possible analogues of chondrules in models where chondrules are produced by impacts on chondrite parent bodies \citep{has96, bas98, sas98, as99}.

\bigskip
\bigskip
\acknowledgments
SAT would like to thank David Matteson, Ray Luo, Joshua Frieman and Arieh K\"{o}nigl for many useful discussions. We thank Conel Alexander, Christine Floss and two anonymous reviewers for useful comments that substantially improved the manuscript. SAT was supported by NASA Origins of Solar Systems Program grant NNX08AM85G. JMP was supported by NASA Cosmochemistry Program grant NNG05GH79Z. JNC was supported by a NASA Origins of Solar Systems Program grant. RCH was supported by a NASA Planetary Geology and Geophysics grant to JNC.

\end{document}